\documentstyle[12pt]{article}
 \renewcommand{\thefootnote}{\fnsymbol{footnote}}

\def\theequation{\arabic{section}.\arabic{equation}}
\begin{document}
\thispagestyle{empty}

\vspace*{2cm}
\begin{center}
{\Large{\bf Antibrackets and non-Abelian equivariant cohomology}} \\

\vspace*{2cm}
{\large A.Nersessian}\footnote{e-mail: nerses@thsun1.jinr.dubna.su  } \\
\vspace*{0.8cm}
{\it Laboratory of Theoretical Physics,\\
 Joint Institute for Nuclear Research,\\
Dubna, Moscow region, Russia 141980}
   \end{center}
\bigskip
\bigskip
\begin{abstract}
The Weil algebra of a semisimple Lie group and an exterior algebra
of a symplectic manifold possess antibrackets.
 They are applied to formulate the models of non--abelian equivariant
cohomologies.
\end{abstract}
\setcounter{page}0
\renewcommand{\thefootnote}{\arabic{footnote}}
\setcounter{footnote}0
\newpage
\section{Introduction}

The Batalin--Vilkovisky formalism (BV-formalism)\cite{bv}
 is the most adequate and powerful method for quantizing gauge fields;
at the same time it is
mathematically the most unusual formalism: its basic structure, antibracket
(see Appendix A), is rather an exotic object.

Study of the BV geometry has shown that  it is actually based on first
principles of the theory of integration on supermanifolds, generalization
of the Stokes theorem to pseudointegral (pseudodifferential) forms (see
\cite{schwarz} and references therein), therefore the antibracket should
play a fundamental role in (super)geometry.

Recently, in field theory, great interest has been displayed in equivariant
cohomologies\footnote{$G$--equivariant cohomology of
$G$- manifold $(M,G)$ is
called the $G$--invariant cohomology of a quotient space $M/G$
\cite{atiah,kalkman}.}, mainly,
in view of the application of the
Duistermaat--Heckman localization formulae \cite{DH}
 connected w
ith $S^1$-equivariant cohomology \cite{atiah},
  to the calculation of path
integrals of topological field theories (see \cite{niemi} and references
therein).
In Ref. \cite{os}, the action of the $4d$ topological Yang--Mills theory was
interpreted in terms of non--Abelian equivariant cohomology, whereas Atiah and
Jeffrey
interpreted its partition function as an Euler regularized equivariant
class \cite{at}. Witten
proposed a generalization of the Duistermaat--Heckman formula to non--Abelian
equivariant cohomologies and used it for calculating the partition function of
the Yang--Mills $2d$ topological theory \cite{witten}.
 This work stimulated a more
detailed study of non--Abelian equivariant cohomologies and search of related
localization formulae \cite{na}.

Non--Abelian equivariant cohomologies are described by a number of
equivariant  models. Specifically, in Refs.\cite{at}, use was made of
the Weil model having a natural geometric interpretation.
The Cartan model convenient for treating
the localization formulae was exploited in Refs.\cite{witten,na}. The authors
of
Ref.\cite{os} introduced the so--called BRST model adapted for
field--theoretical
problems. Kalkman put forward a parametric model of equivariant cohomology
comprising
the Weil and
BRST models
(the latter is naturally reduced to the Cartan model) \cite{kalkman}.

In this note, we construct antibrackets on the basis of Weil algebra $W(g)$
of a Lie semisimple group $G$ and the exterior algebra $\Lambda M$ of a
symplectic $G$-manifold $(M,\omega, G)$, with respect to which the operators
of the exterior  differentiation,
contraction and Lie derivative, are (anti)Hamiltonian. Then we will be able
to formulate the Weil and Cartan models for non--Abelian equivariant cohomology
as well as  a
modified Kalkman model (further, the BRST and Cartan model) in an
(anti)Hamiltonian manner.

The formulation suggested for non--Abelian equivariant cohomology is not only a
convenient tool
for describing equivariant cohomology, rather, it makes richer the theory of
equivariant cohomology and topological
field theories dressing them with the apparatus of BV--formalism and
antisymplectic (super)geometry.

Note that the description of $S^1$-equivariant cohomology
 in terms of antibrackets \cite{jetp} allowed one
to connect thats with an interesting class of supersymmetric mechanics
and to construct generalizations of $S^1$-equivariant characteristic classes.
\setcounter{equation}0
\section{Antibrackets on $\Lambda M$ and $W(g)$.}

Let us construct antibrackets on $\Lambda M$ and $W(g)$ with respect to
which the  operators of differentiation, contraction with generators of the
 $G$-action and the Lie derivative are (anti)Hamiltonian.

Let $(M,\omega,G)$ be a compact symplectic $G$--invariant manifold;
$G$ be a  semisimple Lie
group; $g$ be the Lie algebra of group $G$; $I_a(x)$, $a=1,\ldots{\dim  G}$  be
 a  generators of the
symplectic $G$-action on $(M,\omega)$:
$\omega^{-1}(dI_{a}, dI_{b})= f_{ab}^{d}I_{d}$, where
  $f^{a}_{bd} $  are structure constants of the Lie algebra $g$.

Let $\Lambda M$ be  an (${\bf Z}$--graded) exterior algebra of
$(M, \omega, G)$, and $(x^i,\theta^i)$ be its
local coordinates ($x^i$ are local coordinates of $M$;
$\theta^i$ are the corresponding basic 1--forms:
$\theta^i\leftrightarrow dx^i$).
The ${\bf Z}$--grading on $\Lambda M$ is given by the conditions
 $deg\; x^i =0,\;\;deg\; \theta^i=1$ .

Then, the antibracket on $\Lambda M$ is defined by the expression
\begin{equation}\label{bxt}   \{f,g\}_{\Lambda}=
 \omega^{ij}\left(\frac{\partial f}{\partial x^i}
\frac{\partial _lg}{\partial \theta _j}-\frac{\partial _rf}{
\partial \theta _i}\frac{\partial g}{\partial x^j}\right)
+ \frac{\partial_r f}{\partial \theta^i}
(\theta ^k{\partial \omega ^{ij}}/{\partial x^k})\frac{\partial _lg}
{\partial \theta _j}
\end{equation}
where $\omega^{ij}\omega_{jk}=\delta^i_k$,  $\omega_{ij}=\omega
 ({\partial}/{\partial x^i},{\partial}/{\partial x^j})$.

The functions
\begin{equation}
I_a(x), \quad Q_{a}=
\frac{\partial I_a}{\partial x^i }\theta^i,\quad
D_0=-\frac 12\theta ^i\omega _{ij}\theta ^j ,
\label{iqd}\end{equation}
 define on $\Lambda M$ the (anti)Hamiltonian vector
fields corresponding, respectively, to the contraction of
differential forms with the $\omega^{-1}(dI_a,\;)$,
 to the Lie derivative along $\omega^{-1}(dI_a,\;)$
and to the external differentiation \cite{jetp}.\\

Now consider the (${\bf Z}$--graded) Weil algebra
$ W(g)=S(g^*)\otimes\Lambda(g^*)$ of the Lie group $G$. Here
$S(g^*)$ is a symmetric algebra of  polynomials on the algebra
$g^*$, dual to $g$, with (commuting) coordinates $\phi^{a}$;
 $\Lambda({g}^*)$ is an external algebra on $g^*$ with (anticommuting)
coordinates $c^a$. The ${\bf Z}$--grading on $W(g)$ is given by the
conditions $deg\phi^a =2,\;\;deg c^a =1$
(this grading is chosen due to correspondence of $c^a$ and $\phi^a$ ,
respectively, to the  the connection
1-form on the principal $G$--fiber bundle and to the its curvature).

The antibrackets on $W(g)$ can be defined by the formula
\begin{equation}
\{f , g\}_{W} = g^{ab}\left(\frac{\partial f}{\partial \phi^a}
\frac{\partial _lg}{\partial c^b}-
 \frac{\partial_r f}{\partial c^a}\frac{\partial g}{\partial \phi^b}\right),
\label{b}\end{equation}
where $ g^{ad}g_{db}=\delta^a_b$, and $g_{ab}$ is
the Cartan--Killing metric of algebra $g$.

The functions
\begin{equation}
\phi_{a}= g_{ab}\phi^{b} , \quad q_{a}= f^{d}_{ba}\phi_d c^b, \quad D=
\frac{1}{2}g_{ab}\phi^a\phi^b -
 \frac{1}{2}f^{d}_{ab}\phi_d c^a c^b.
\label{dpm}\end{equation}
act on $W(g)$ as follows:
  \begin{eqnarray}
&\{ \phi_b, c^a\}_{1} =\delta^a_b ,\quad\quad \{\phi_b, \phi^a\}_{1} =0; \quad
\{ q_b, c^a\}_{1} =f^a_{db} c^d,\quad\quad \{q_b,\phi^a\}_{1}= f^a_{db}\phi^d
.&\\
 &\{ D, c^a\}_{1} =
\phi^{a} - \frac{1}{2}f^{a}_{bd}c^bc^d \quad \{D, \phi^a\}_{1}
=-f^{a}_{bd}c^b\phi^d ,&
\label{cw}\end{eqnarray}
They define, respectively, the contraction of the
connection 1--form with generators of the $G$-action  on
$W(g)$, co--adjoint action of $G$ on $W(g)$ and the Weil differential.

Equipping the metric $ g_{ab} $ with the  grading
 $deg\;\; g_{ab} =-2$ , we obtain
\begin{equation}
deg\;\{\;\;,\;\;\}_{\alpha}=-1 ;\quad
deg\;\;\left (I^{\alpha}, \;\;Q^{\alpha},\;\;D^{\alpha}\right)=(0, 1, 2),
\end{equation}
 where
\begin{equation}
\{\;\;,\;\;\}_{\alpha}=(\{\;\;,\;\;\}_{\Lambda},\;\{\;\;,\;\;\}_{W}),\quad
\left(I^{\alpha}_a, Q^{\alpha}_a, D^{\alpha}\right)=
\left( (I_a, Q_a, D_0), (\phi_a, q_a, D)\right).
 \nonumber\end{equation}
The sets (\ref{iqd}) and (\ref{dpm}) form,
with respect to a corresponding  antibrackets, the superalgebra
\begin{eqnarray}
 && \{I^{\alpha}_{a}, Q^{\alpha}_{b}\}_{\alpha}
=f_{ab}^{d}I^{\alpha}_{d},\quad
\{Q^{\alpha}_{a}, Q^{\alpha}_{b}\}_{\alpha} =f_{ab}^{d}Q^{\alpha}_{d} ,\quad
 \{I^{\alpha}_{a}, D^{\alpha}\}_{\alpha} =Q^{\alpha}_{a},\quad\\
 &&\{D^{\alpha} , D^{\alpha}\}_{\alpha} =0,
    \quad\{Q^{\alpha}_{a}, D^{\alpha}\}_{\alpha} =0,
 \quad \{I^{\alpha}_{a}, I^{\alpha}_{b}\}_{\alpha}=0 ,\nonumber
\label{5} \end{eqnarray}

\section{Models for equivariant cohomology}

Now, we are able to construct a models for equivariant cohomology.\\

{\bf The Weil model.} On the space ${\cal A}=\Lambda (M)\otimes W(g)$,
the antibracket
   \begin{equation}
\{f, g\}_{\cal A}=\{f,g\}_{\Lambda}+\{f,g\}_{W}  \label{wb}\end{equation}
is defined.

 The functions
 \begin{equation}  {\cal I}_a=(I_a +\phi_a),\quad
{\cal Q}_a=(Q_a +q_a),\quad {\cal D}=(D_{0}+D ),\quad .
  \label{whole} \end{equation}
form on  $({\cal A}, \{\;\;,\;\;\}_{\cal A})$  the superalgebra
 (\ref{5}).
 They generate anti--Hamiltonian vector fields
corresponding to the contraction with generators of the $G$-action , Lie
derivative along them and to the total differential,
respectively.

The $G$ equivariant cohomology of the manifold $M$ is defined as a
subspace of ${\cal A}$ whose elements
commute with all ${\cal I}_a$ and ${\cal Q}_a$
with respect to the antibrackets (\ref{wb}).
It is a cohomology of the operator, generating by the function
${\cal D}$  \cite{atiah,kalkman}.\\

{\bf The Kalkman-like model.}
 Let us deform the Weil model performing the
(anti)ca\-no\-ni\-cal transformation preserving the ${\bf Z}$--grading and
generated by    the function
 \begin{equation} \Psi =-tc^a I_{a} (x) , \;\;deg\;\;\Psi =1,
 \end{equation}
where $t$ is a parameter.

The antibracket (\ref{wb}) is invariant under these transformations by
definition, and the Hamiltonians (\ref{whole}) are transformed to the following
ones:
\begin{eqnarray} {\cal I}^t_a&=&
{\cal I}_a - t\{{\cal I}_a , \Psi\}_{1} +\frac{t^2}{2!} \{\{{\cal I}_a ,
\Psi\}_{1},\Psi\}_{1} +...=\nonumber\\
&=& \phi_a +(1-t)I_a \label{i}\\
{\cal
 Q}^t_a&=& {\cal Q}_a  - t\{{\cal Q}_a , \Psi\}_{1} +\frac{t^2}{2!} \{\{{\cal
Q}_a , \Psi\}_{1},\Psi\}_{1} +...=\nonumber\\
&=&{\cal Q}_a .\label{q} \\
{\cal D}^t &= &{\cal D} - t\{{\cal D} ,
\Psi\}_{1} +\frac{t^2}{2!} \{\{{\cal D} , \Psi\}_{1},\Psi\}_{1}
+...=\nonumber\\
&=&  {\cal D} - t\phi^aI_a +\frac{t^2}{2}g^{ab}I_a I_b
 +tc^aQ_a +\frac{t(1-t)}{2}f^c_{ab}c^a c^bI_c\label{brst2}
\end{eqnarray}
They form the superalgebra (\ref{5}) with respect to
(\ref{wb}) at any values of the parameter $t$.
The operators defined by the functions (\ref{i}) and (\ref{q})
coincide with the corresponding operators of the Kalkman parametric model
for equivariant cohomology \cite{kalkman}.
The equivariant cohomology in this model is defined like in the Weil model.

The differential given by the function (\ref{brst2})
differs from that proposed by Kalkman:
\begin{equation}
{\cal D}^{K}=\{{\cal D},\;\;\}_{\cal A}
- t(\phi^a-tI_ag^{ab})\{I_a,\;\;\}_{\cal A} +
\frac{t(1-t)}{2}f^c_{ab}c^ac^b\{I_a,\;\;\}_{\cal A}  -
-tc^a \{ Q_a,\;\;\}_{\cal A},
\label{dkalkman}\end{equation}
that is not Hamiltonian with respect to (\ref{wb}).\\

{\bf BRST-like   model.}
 At $t = 1$ the functions (\ref{i})--(\ref{brst2})
assume the form
\begin{equation} {\cal I}^1_a= \phi_a ,\quad
{\cal Q}^1_a={\cal Q}_a ,\quad {\cal D}^1 =
 {\cal D} -\phi^aI_a +\frac{1}{2}g^{ab}I_a I_b +c^aQ_a. \label{i3}
\end{equation}
Generators generated by the first two functions coincide with the
corresponding generators of the standard BRST model
(arising in the  $4d$ topological Yang-- Mills theory \cite{os})
to which the Kalkman model is reduced at $t = 1$.

The function ${\cal D}^1$ generates the differential
different from the standard BRST one
 \begin{equation}
{\hat{\cal D}}^{BRST} = \{{\cal D},\;\;\}_{\cal A} -
\phi^a\{I_a,\;\;\}_{\cal A}  + c^a\{Q_a,\;\;\}_{\cal A}.
\label{brst1}  \end{equation}
The equivariant cohomology  in the BRST model belongs to the space
$\Lambda (M)\otimes S({g^*})$ on which by the expression
 (\ref{bxt}) the degenerate  antibracket is defined.\\

{\bf  Cartan-like models.}Restriction of the BRST model to the space
 $\Lambda (M)\otimes S({g^*})$ results in the Cartan model.
It is defined by the functions
$(D_0\pm I_\phi)\equiv D_0\pm\phi^a I_{a},\quad
Q_\phi\equiv\phi^a Q_{a},$ forming the superalgebra
\begin{eqnarray}
&\{I_{\phi}\pm D_0 , I_{\phi} \pm D_0 \}_{1} =\pm 2Q_{\phi},\;\;
\{I_{\phi}+ D_0, I_{\phi} - D_0\}_{1} = 0&\\
&\{I_{\phi}\pm D_0 , Q_{\phi} \}_{1} =
\{Q_{\phi}, Q_{\phi} \}_{1} = 0 .&
 \nonumber \label{sualg2} \end{eqnarray}
The function $D_0 -I_{\phi}$
determines the differential in the Cartan model nilpotent
on the space of $G$--invariant elements of the space $\Lambda M$.
 Therefore, the $G$-equivariant cohomology of
the manifold $\Lambda M$ is an element of $S(g^*)\otimes\Lambda M$
commuting with $D_0 -I_{\phi}$ with respect to the
antibracket (\ref{bxt}) \cite{kalkman}.\\

The restriction of the differential of the standard BRST model to
$S(g^*)\otimes\Lambda M$  gives rise to that of the standard Cartan model.
 The limitation
of the differential generated by the function (\ref{brst2})
at $t = 1$ results in  a different differential
 \begin{equation}
{\cal D}_C^1 =  D_0 +\frac 12 g^{ab}(\phi_a -I_a)(\phi_b -I_b) :\;\;
\{ {\cal D}_C^1,{\cal D}_C^1\}_1 = (\phi^a -I^a) Q_a.
\end{equation}

{\bf Acknowledgements.}
 The author is thankful to  A. Hietam\"aki for useful
comments, and to I.A. Batalin and A.Niemi for interest in the work.
 This work was
supported by the Grant No. M2I300 from International Scientific Foundation
and by  the INTAS Grant No. 93--2492)  and has been carry out within the
research program of the
International Center of Fundamental Physics in Moscow.\\
\def\theequation{A\arabic{equation}}
\appendix
\setcounter{equation}0
\vspace*{0.5cm}
{\large\bf Appendix}

\section{ Antibrackets}

The antibracket of the functions $f(z)$, $g(z)$ on the supermanifold
${\cal M}$
 is called the operation
\begin{equation} \label{bodd}
\{f,g\}=\frac{\partial _rf}{\partial z^A}\Omega^{AB}_{1} \frac{\partial
_lg}{\partial z^B},
\end{equation}
(where $r$ and $l$ denote, respectively, the right-- and
left--handed derivatives) obeying the conditions
\begin{eqnarray} &  p(\{ f, g \} )= p(f)+ p(g) + 1 , \quad {\rm (grading
condition)}&  \nonumber \\
 & \{ f, g \} = -(-1)^{(p(f)+1)(p(g)+1)}\{ g, f \}, \quad {\rm (
``antisymmetrysity")}
& \label{anti} \\ &  \{ f,\{ g, h \}\} -(-1)^{(p(f)+1)(p(h)+1)} \{ g,\{ f, h
\}\}= \{\{ f, g \}, h\} .\quad{\rm {(Jacobi \quad id.)}} & \label{bjac}
\end{eqnarray}
With every function $f$, the antibracket associates an operator
(anti--Hamiltonian vector field) of opposite parity
$ {\hat f}=\{f,\;\;\}$, and, in view of the
Jacobi identity (\ref{bjac}), there holds the following relation:
 $$ {\hat {\{ f, g\}}}={\hat f}{\hat g} -
(-1)^{p({\hat f})p({\hat h})}{\hat g}{\hat f} .$$
These fields generate transformations preserving the antibracket
(anticanonical transformations).

On the $(n.n)$--dimensional supermanifold, antibrackets can be nondegenerate.
Then they can be associated with the antisymplectic  structure
\begin{equation} \label{symp} \Omega=dz^A\Omega _{AB}dz^B , \quad
d\Omega =0,\quad \Omega _{AB}\Omega^{BC}_{1}=\delta _A^C \end{equation}
Locally, the antibrackets are reducible to the canonical form \cite{leites}
\begin{equation}\label{bcan}
 \Omega^{{\rm can}}=\sum_{i=1}^n dx^{i}\wedge d\eta_{i}
,\quad   \{f,g\}^{{\rm can}}=\sum_{i=1}^n\left( \frac{\partial _rf}{%
\partial x^i}\frac{\partial _lg}{\partial \eta _i}-\frac{\partial _rf}{%
\partial \eta _i}\frac{\partial _lg}{\partial x^i}\right),
\end{equation}
where  $ p(\eta_i) =  p(x_i) +1$.
On the space $\Lambda_* M$ of polyvector fields of the
manifold $M$, the canonical antibracket can be defined globally; in this case
 $x^i$ are local coordinates of $M$ and $\eta_i$
are basis vector fields:  $\eta_i\leftrightarrow \tilde
 \frac{\partial}{\partial x^i}$ .

Antisymplectic structures on $\Lambda M$ and $W(g)$
corresponding to the antibrackets  (\ref{bxt}) and (\ref{b}) are,
respectively, given by the expressions
\begin{equation}\label{aL}
\Omega_{\Lambda}= \omega_{ij}dx^{i}\wedge d\theta^{j}
 +\frac 12\omega_{ij,k}\theta^k dx^{i}\wedge dx^{j},
\quad\Omega_{W}=g_{ab}d\phi^{a}\wedge dc^{b} .
\end{equation}

 \end{document}